# Pyrochlore Oxide $Hg_2Os_2O_7$ on Verge of Metal–Insulator Boundary

Kota Kataoka[1], Daigorou Hirai[1], Akihiro Koda[2], Ryosuke Kadono[2], Takashi Honda[2] and Zenji Hiroi[1]

[1]*Institute for Solid State Physics, University of Tokyo, Kashiwa, Chiba 277-8581, Japan*
[2]*Institute of Materials Structure Science, High Energy Accelerator Research Organization (KEK-IMSS), Tsukuba, Ibaraki 305-0801, Japan*

Semimetallic osmium pyrochlore oxide $Cd_2Os_2O_7$ undergoes a magnetic transition to an all-in–all-out (AIAO)-type order at 227 K, followed by a crossover to an AIAO insulator at around 210 K. Here, we studied the isostructural and isoelectronic compound $Hg_2Os_2O_7$ through thermodynamic measurements, $\mu$SR spectroscopy and neutron diffraction experiments. A similar magnetic transition, probably to an AIAO-type order, was observed at 88 K, while the resistivity showed a decrease at the transition and remained metallic down to 2 K. Thus, the ground state of $Hg_2Os_2O_7$ is most likely an AIAO semimetal, which is analogous to the intermediate-temperature state of $Cd_2Os_2O_7$. $Hg_2Os_2O_7$ exists on the verge of the metal–insulator boundary on the metal side and provides an excellent platform for studying the electronic instability of $5d$ electrons with moderate electron correlations and strong spin–orbit interactions.

## 1. Introduction

The study of materials containing heavy transition metals is at the forefront of solid state chemistry and condensed matter physics. In comparison to $3d$ compounds, $5d$ compounds exhibit a relatively strong spin–orbit interaction (SOI), which frequently plays a critical role in electronic properties, just as electron correlation does in $3d$ compounds [1]. We are particularly interested in the properties of $5d$ electrons near the metal–insulator (MI) boundary, which must be distinguished from the Mott–Hubbard (MH) physics based on electron correlations in $3d$ compounds.

The pyrochlore oxides $Dy_2Ir_2O_7$ and other iridates with various lanthanide elements [2-5] and $Cd_2Os_2O_7$ [6-9] have been the most extensively studied $5d$ compounds in this context. Both have the same cubic $Fd$–$3m$ structure but have essentially different electron configurations ($Ir^{4+}$: $5d^5$, $Os^{5+}$: $5d^3$) and band structures [10-13]: $Dy_2Ir_2O_7$ has a quadratic band touching at the $\Gamma$ point at the Fermi energy ($E_F$), whereas $Cd_2Os_2O_7$ is a compensated semimetal with an even number of electrons and holes surrounding the low-symmetry points in the Brillouin zone. Despite their distinct electronic structures, they share numerous similarities: upon cooling, both exhibit metal–insulator transitions (MITs) accompanied by all-in–all-out (AIAO) type magnetic orders in the pyrochlore lattice made of $5d$ atoms [9,14-17]. Additionally, magnetic domain walls (MDWs) associated with the AIAO-type order possess robust ferromagnetic moments and retain their metallic conductivity even when the bulk of both compounds becomes insulating [18-22].

The MIT in $Cd_2Os_2O_7$ has been proposed to be a magnetic Lifshitz transition [20,23,24]: the AIAO-type order induces a shift or flattening of the $5d$ semimetallic bands, thereby removing their overlap at $E_F$ and forming an indirect gap. The driving force has been ascribed to electron correlations as well as SOIs [15,25,26]. Numerous recent experiments provide evidence for this mechanism, as well as additional intriguing phenomena such as strong spin–phonon couplings and coupled spin–charge–phonon fluctuations [27-31].

Sohn et al. observed that a charge gap did not open at $T_N$ but did so below the crossover temperature $T^*$ of ~210 K in their optical measurements on $Cd_2Os_2O_7$ [23]. Thus, an intermediate metallic AIAO state exists within a narrow temperature window between $T_N$ and $T^*$; $T_N$ and $T^*$ need not coincide. This is because, unlike the first-order MH transition, the overlap between the hole and electron bands in the semimetallic band structure can be gradually reduced below $T_N$ in a second-order manner and eventually vanishes at $T^*$ when sufficient magnetic correlations develop. Note that this MIT should appear as a crossover, because the conductivity near $T^*$ is dominated by a large number of thermally excited carriers across a vanishingly small band gap; such a Lifshitz transition with a topological change of the Fermi surface but no symmetry breaking becomes a phase transition only at zero temperature.

To gain better understanding of the mechanism of $Cd_2Os_2O_7$'s MIT, we studied another metallic pyrochlore oxide, $Hg_2Os_2O_7$, which has the same crystal structure and electron configuration as $Cd_2Os_2O_7$. $Hg_2Os_2O_7$ was unknown in 1983 when Subramanian et al. published their review paper on pyrochlore oxides [32], and Reading et al. synthesised it in polycrystalline form for the first time in 2002 [33]. They demonstrated using neutron diffraction (ND) experiments that the compound crystallised in a cubic pyrochlore structure over a wide temperature range of 12–300 K. Moreover, they discovered an anomaly in the Curie–Weiss (CW) magnetic susceptibility and a kink in the resistivity at 88 K, followed by a more metallic behaviour down to 4 K. They did, however, suggest the absence of long-range order (LRO) due to frustration because they did not detect magnetic reflections in their powder ND experiments. On the other hand, Tachibana et al. reported a nearly temperature-independent, Pauli paramagnetic-like magnetic susceptibility with a small anomaly in the zero-field cooling curve at 88 K and an upturn in the field-cooling curve at the same temperature [34]. They did, however, suggested the absence of LRO because no anomaly in heat capacity was observed at this temperature. As a result, the fundamental properties of $Hg_2Os_2O_7$ have remained obscure and somewhat controversial, particularly regarding the presence of a phase transition at 88 K.

Here, we prepared single crystals of $Hg_2Os_2O_7$ and conducted a detailed analysis of the sample dependences of magnetic susceptibility, resistivity, and heat capacity. We discovered that LRO did appear at $T_N$ = 88 K in potentially high-quality crystals with metallic conductivity preserved down to 2 K. $\mu$SR experiments established that the LRO was magnetic in origin. In comparison to $Cd_2Os_2O_7$, it appears that $Hg_2Os_2O_7$ takes a semimetallic state with an AIAO-type order in the ground state, which may correspond to the intermediate phase of $Cd_2Os_2O_7$.

## 2. Experimental

A polycrystalline sample was prepared by heating a mixture of HgO and Os to 653 K for 150 hours in a sealed silica tube with 12 mm diameter and 150 mm length. Additional oxygen was supplied by the thermal decomposition of AgO, which was placed separately in the silica tube. The concentrations of HgO, Os, and AgO were chosen to give the stoichiometry of $Hg_2Os_2O_7$ when AgO was completely reduced to Ag, as actually observed after the reaction. Special care was required to avoid the silica tube exploding due to the volatile nature of Hg and Os, as well as to avoid the production of highly toxic $OsO_4$.

Two methods were used to grow single crystals of $Hg_2Os_2O_7$. An aggregate of small crystals was occasionally obtained by heating the same starting material as above to a higher temperature of 773 K for 100 hours in a box furnace [crystal A, figure 1(a)]. The chemical vapour transport method was used to grow small isolated octahedral crystals such as those shown in figure 1(b) (crystals B; B1 and B2). The transport agent was $HgCl_2$, and the oxygen source was $NaClO_3$. A silica tube with a diameter of 15 mm and a length of 350 mm was heated for 72 hours at a temperature gradient of 773–673 K, containing a pellet of a mixture of Os, HgO and $HgCl_2$, as well as $NaClO_3$ placed separately from the pellet. Typically, the initial composition of Os : HgO : $HgCl_2$ : $NaClO_3$ was 2 : 1 : 0.5 : 2.

The thus-obtained samples were characterised using powder X-ray diffraction (XRD) using Cu–K$\alpha$ radiation in a diffractometer (RINT-2000) and chemical analysis in an energy-dispersive X-ray (EDX) analyser (JSM-IT100). A Quantum Design PPMS was used to measure resistivity, Hall effect, and heat capacity, and a Quantum Design MPMS-3 was used to measure magnetisation. The polycrystalline sample and the two types of crystals were used to investigate sample dependence.

$\mu$SR experiments were conducted at TRIUMF using the LAMPF spectrometer on the M20 beamline, and ND experiments were conducted at the Material and Life Science Experimental Facility (MLF) of the Japan Proton Accelerator Research Complex (J−PARC) using the neutron total scattering spectrometer NOVA (BL21). Both measurements were performed on polycrystalline samples. Fullprof software was used to simulate ND patterns [35].

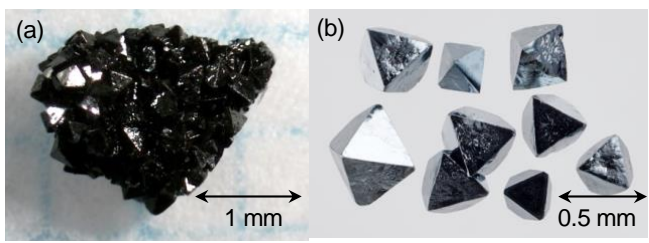

**Figure 1.** Photographs of typical $Hg_2Os_2O_7$ single crystals: (a) aggregate crystal A; (b) small single crystals B.

## 3. Results
### 3.1 Sample characterisations

Figure 2 shows the powder XRD patterns at room temperature for a polycrystalline sample and two crushed crystals A and B1. They indicate that the samples were monophasic and free from impurities. All the peaks were indexed to cubic unit cells of the pyrochlore type structure, which have lattice constants of 10.2488, 10.248 and 10.2449 Å for the polycrystalline sample, crystal A, and crystal B1, respectively, which are in reasonably good agreement with the previous ND value of 10.24681 Å [33]. Note that the relative intensities of the diffraction peaks vary between samples, which may be partly explained by the preferential orientation along the [111] direction, as indicated by the comparison to the simulated XRD pattern.

The chemical compositions of Hg and Os were determined by EDX analysis. With a content of oxygen of 7, the compositions of (Hg, Os) were (2.07, 1.97), (2.05, 1.98), and (2.07, 1.97) for crystals A, B1 and B2, respectively. The metal compositions of the crystals differed slightly, within the experimental error of about 0.01–0.02 per formula unit. The Hg composition greater than 2 may be due to experimental error caused by the difficulty in selecting a standard sample (HgTe was used), because such an excess of metals has never been observed in pyrochlore oxides; the densely packed structure lacks interstitial sites [32]. We were unable to determine the oxygen content due to experimental difficulties with iodometry and the ease with which $Hg_2Os_2O_7$ decomposes into toxic compounds. As a result, our chemical analyses were unable to identify a possible difference in the compound's stoichiometry.

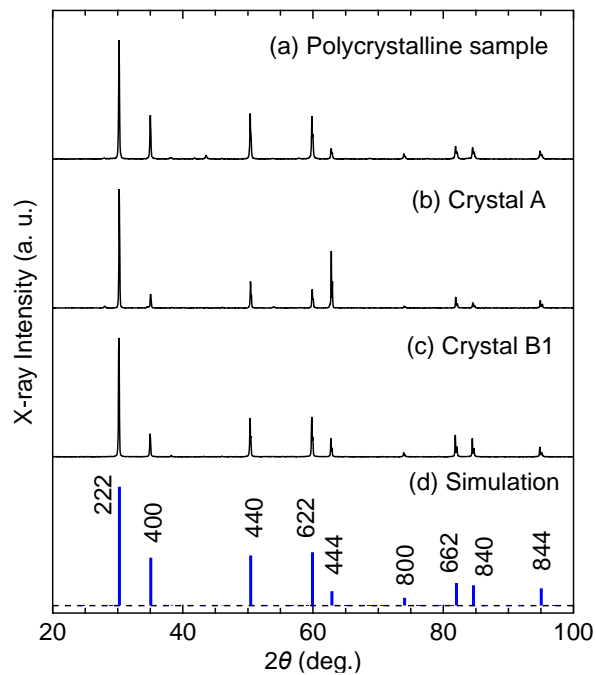

**Figure 2.** Powder XRD patterns of $Hg_2Os_2O_7$ (a) polycrystalline sample, (b) crushed crystal A, and (c) crushed crystal B1. A corresponding simulated pattern based on the structural parameters provided in a previous study [33] is shown in (d).

### 3.2 Fundamental properties and sample dependences

The resistivity and magnetic susceptibility of the three crystals are shown in figure 3. Resistivities were commonly

~1 mΩ cm at 300 K, which is a typical value for a metal near the MI boundary [36]. When crystal A is cooled to 86 K, the resistivity exhibits a kink, as indicated by the peak in its derivative curve, followed by a more metallic temperature dependence toward a large residual value. Similarly, the magnetic susceptibility of crystal A exhibits an anomaly at 88 K, below which the heating curve after zero-field cooling (ZFC) and the subsequent cooling curve in the same magnetic field (FC) begin to diverge (we call this temperature $T_N$). In comparison, crystals B1 and B2 exhibit comparable flattest resistivity curves with only minor anomalies at ~65 K. At comparable temperatures, their magnetic susceptibilities exhibit similar changes to those of crystal A, with the exception of the larger separations between the ZFC and FC curves. The temperature dependences of resistivity and magnetic susceptibility of crystal A are similar to those of the polycrystalline sample reported by Reading et al. [33].

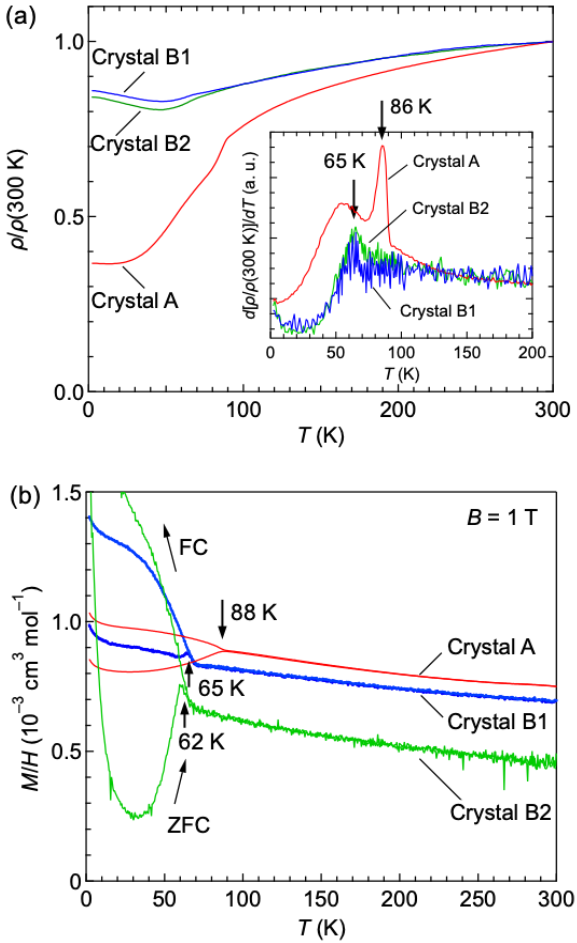

**Figure 3.** Temperature dependence of (a) resistivity normalised to 300 K and (b) magnetic susceptibility for three samples: aggregate crystal A and two crystals B1 and B2. The inset of (a) illustrates peaks at 86 and 65 K in the temperature derivatives of the resistivity curves of crystal A and crystals B, respectively. Magnetic susceptibility was measured in a 1 T magnetic field upon heating after zero-field cooling (ZFC) and subsequently upon cooling in the same magnetic field (FC).

Heat capacity data for crystal A and several crystals from the same sample batch as crystals B1 and B2 (crystals B) are shown in figure 4. At approximately 80 K for crystal A, a hump on the large lattice contribution is observed: it grows below 90 K and merges with the lattice contribution at ~60 K. The hump must correspond to anomalies in resistivity and magnetic susceptibility observed at $T_N$, indicating a thermodynamic phase transition in the bulk. However, the transition is significantly broadened, possibly due to certain crystallographic imperfections: one section of the crystal has a critical temperature of 88 K, while the remaining sections have lower critical temperatures. Although the transition entropy is difficult to calculate, it is approximately 0.67 J K$^{-1}$ mol$^{-1}$, which is significantly less than the maximal entropy for the localised spin 3/2 of $d^3$ electrons of $2R\ln 4$ = 23.0 J K$^{-1}$ mol$^{-1}$.

For crystals B, in contrast, only a smooth variation without anomalies is observed, indicating that the phase transition has been widened further due to increased inhomogeneity; heat capacity is frequently more sensitive to inhomogeneity than resistivity or magnetic susceptibility. The lower residual-resistivity ratios (RRRs) and larger separations between the ZFC and FC curves in magnetic susceptibility also demonstrate the poor quality of crystals B, the latter of which has been ascribed to many MDWs trapped by crystalline defects [20,21]. The fact that crystals B1 and B2 have lower transition temperatures may also be due to disorder. The absence of an anomaly in the heat capacity data by Tachibana et al. [34] could be a result of this sample quality issue.

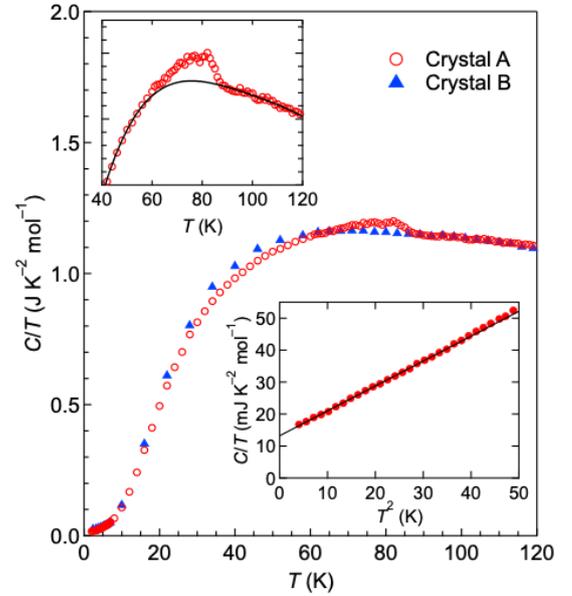

**Figure 4.** Heat capacity divided by temperature $C/T$ of aggregate crystal A and several crystals from a sample batch of crystals B. The upper inset enlarges the data around the transition. The solid line represents a phonon background obtained by fitting the data at 40–50 K and above 90 K to a fourth-order polynomial. The lower inset shows a plot of $C/T$ versus $T^2$ at low temperatures below 7 K, in which a linear fit gives a Sommerfeld coefficient of 13.3 mJ K$^{-2}$ mol$^{-1}$.

The observed sample dependence is not well understood. It could be a minute change in stoichiometry or structural disorder that is difficult to detect experimentally. Indeed, previous ND experiments revealed no structural abnormalities [33]. Even the quality of crystal A is unsatisfactory, as it exhibits a broad transition in heat capacity and a low RRR of 2.7, indicating that the crystal quality must be improved prior to conducting detailed investigation. However, for the purpose of this study, we used crystal A as the highest-quality sample for subsequent experiments.

The heat capacity of crystal A at low temperatures is fitted to the equation $\gamma + \beta T^3$ using a Sommerfeld coefficient $\gamma$ of 13.3 mJ K$^{-2}$ mol$^{-1}$; the $\gamma$ of crystals B was 20 mJ K$^{-2}$ mol$^{-1}$ (not shown) and the previously reported value was 21 mJ K$^{-2}$ mol$^{-1}$ [34]. These values are significantly greater than ~1 mJ K$^{-2}$ mol$^{-1}$ for Cd$_2$Os$_2$O$_7$ [7], which is consistent the metallic ground state of Hg$_2$Os$_2$O$_7$ having a finite density of states (DOS) at $E_F$. The calculated Debye temperature from $\beta$ is 270.2 K, which is lower than the values for Cd$_2$Os$_2$O$_7$ of 463 or 354 K [7].

The Hall coefficient $R_H$ of crystal A at room temperature is negative, as shown in figure 5. It is nearly temperature-independent above 150 K, becomes positive with a peak near $T_N$, and then reverts to a negative value to $-6.7 \times 10^{-4}$ cm$^3$ C$^{-1}$ at 2 K. Due to the fact that the Hall resistivity is linear with the magnetic field up to 7 T, electron carriers with higher mobilities dominate $R_H$ in both the high-temperature compensated semimetallic and the low-temperature ground states. On the other hand, the sign changes and peak formation near $T_N$ are ascribed to competition between electron and hole carriers, which is caused by a modification in the band structure around the magnetic transition.

The $R_H$ of Hg$_2$Os$_2$O$_7$ is compared to that of Cd$_2$Os$_2$O$_7$ in figure 5. The overall temperature dependence of $R_H$ is similar for the two compounds: Cd$_2$Os$_2$O$_7$ has a similar peak around $T_N$, which is too small to see in the figure [20]. This indicates that the semimetallic band structures of the two compounds share some characteristics and modifications. On the other hand, the absolute value for Hg$_2$Os$_2$O$_7$ at 300 K is $1.1 \times 10^{-4}$ cm$^3$ C$^{-1}$, which is significantly less than the value of $7.0 \times 10^{-4}$ cm$^3$ C$^{-1}$ for Cd$_2$Os$_2$O$_7$, indicating that Hg$_2$Os$_2$O$_7$ has a higher carrier density in the semimetallic states. At the lowest temperature, the absolute value of Hg$_2$Os$_2$O$_7$ is three orders of magnitude smaller than that of Cd$_2$Os$_2$O$_7$, indicating metallic and insulating ground states, respectively.

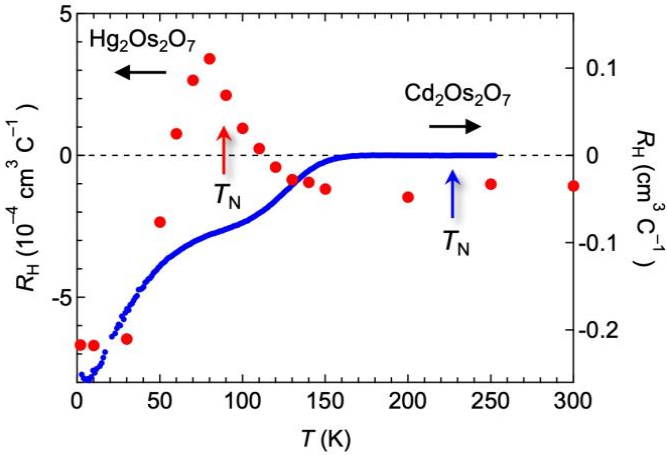

**Figure 5.** Hall coefficient $R_H$ measured for crystal A, which was obtained from the slope of Hall voltage versus magnetic field curves up to ±10 T at each temperature. The $R_H$ of Cd$_2$Os$_2$O$_7$ is also plotted on the right ordinate [20].

*3.3 Magnetic long-range order*

It is now clear from the above-mentioned thermodynamic data that Hg$_2$Os$_2$O$_7$ undergoes a phase transition at $T_N$ = 88 K. Given that a well-defined anomaly in magnetic susceptibility is observed, it must be a transition to a magnetic LRO. We conducted $\mu$SR experiments on a polycrystalline sample to obtain more direct evidence. As with Cd$_2$Os$_2$O$_7$ [37], the $\mu$SR time spectra in figure 6 exhibit an oscillatory component due to an internal magnetic field at 2.3 K. At 80 K, the oscillatory component persists, becomes obscured at 85 K, and finally vanishes entirely at 102 K. Thus, it is evident that a magnetic LRO establishes itself at $T_N$. However, the coexistence of a non-oscillatory component even at the lowest temperature suggests that the LRO is associated with an inhomogeneous distribution of the internal magnetic field akin to that of a spin-density-wave (SDW). The transition temperature was estimated to be 91(1) K based on the temperature dependence of the oscillatory components obtained by fitting the data to a two-component model; details of the analysis will be published elsewhere [38].

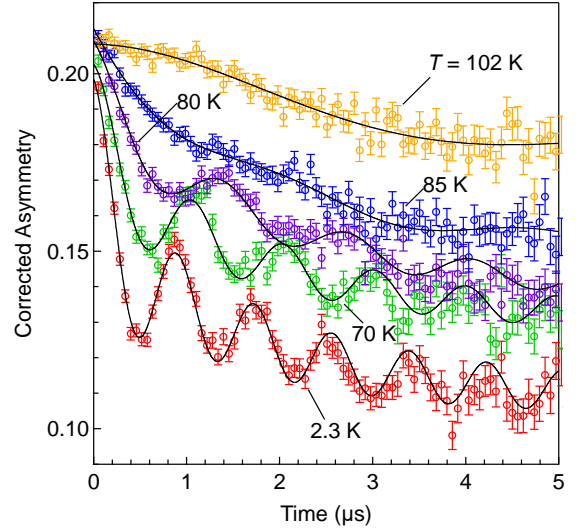

**Figure 6.** Temperature evolution of $\mu$SR time spectra in the absence of a magnetic field, where $\mu$–$e$ decay asymmetry, which is proportional to muon polarisation, is plotted. Each dataset contains a solid curve that represents a fit to a two-component model [38]. Below ~85 K, an oscillatory component originating from the LRO's internal magnetic field grows.

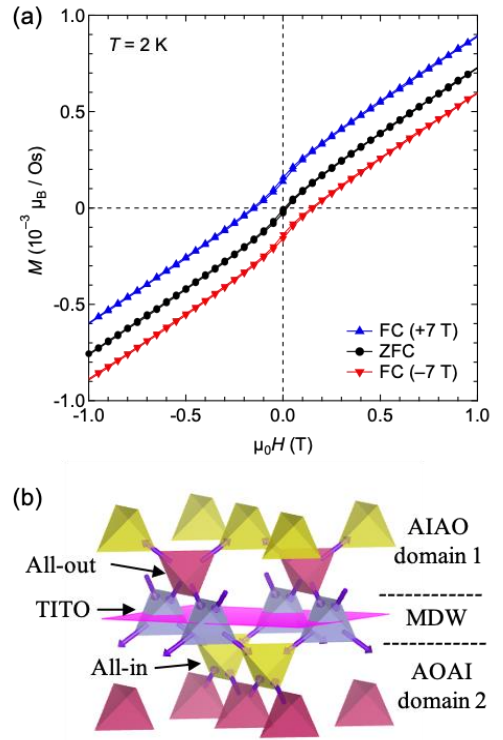

**Figure 7.** (a) $M$–$H$ curves at 2 K following ZFC and FC in magnetic fields of ±7 T. Along with a dominant linear component, each curve contains a minor contribution that saturates at low magnetic fields and must originate from a trace of a ferromagnetic impurity phase. After FC, a remnant magnetisation of ~$2 \times 10^{-4} \mu_B$ per Os appears, which changes sign with field inversion at the FC process. (b) Schematic representation of an AIAO-type magnetic domain wall (MDW). Domain 1 has an all-in–all-out (AIAO) spin arrangement,

whereas domain 2 has an all-out–all-in (AOAI) spin arrangement. At the interface between them, an MDW composed of tetrahedra with a two-in–two-out (TITO) spin arrangement is generated.

The similarity in magnetic susceptibility between the two compounds suggests that $Hg_2Os_2O_7$ also exhibits an AIAO-type LRO. Moreover, similar to $Cd_2Os_2O_7$, we observed robust ferromagnetism in $Hg_2Os_2O_7$ (figure 7). It has been ascribed to MDWs carrying ferromagnetic moments embedded in the matrix of the AIAO-type antiferromagnetic order in the case of $Cd_2Os_2O_7$ [20,21]. As illustrated schematically in figure 7(b), an MDW composed of tetrahedra with the two-in–two-out (TITO) spin configuration is generated at the interface between the two antiphase domain types of the AIAO-type order. Since TITO tetrahedra carry uncompensated moments perpendicular to the interface, the MDW should have a ferromagnetic moment parasitic on the AIAO-type order in the matrix. Notably, the ferromagnetic moment can be reversed only when the MDW moves to the next layer of tetrahedra, which is only possible at elevated temperatures and with a soft magnetic order.

As shown in figure 7(a), when crystal A is measured at 2 K after zero-field cooling from above $T_N$, it exhibits a symmetric M–H curve around the origin. In comparison, the M–H curve shifts upward with a positive remnant magnetisation $M_r$ at zero field after field cooling in a +7 T magnetic field and downward with a negative $M_r$ after field-cooling in a reversed field of –7 T. This was precisely the case for $Cd_2Os_2O_7$ [21]. Notably, the ferromagnetic component is extremely robust: it does not flip in the presence of a reverse magnetic field of ±7 T (only M–H curves between +1 and –1 T are shown in the figure) because the movement of MDWs caused by flipping spins with large Ising anisotropy is disturbed at low temperatures. $M_r$ magnitudes of approximately $2 \times 10^{-4} \mu_B$ per Os are twice that of a typical $Cd_2Os_2O_7$ crystal, implying a higher density of defects trapping MDWs in the $Hg_2Os_2O_7$ crystal. As a result, it is plausible that a similar AIAO-type LRO occurs in $Hg_2Os_2O_7$.

We conducted powder ND experiments at 300, 100 and 50 K to investigate the possibility of an AIAO-type order (figure 8). All of the diffraction profiles are consistent with the $Fd\bar{3}m$ structure previously observed by ND [39]. Magnetic contributions are calculated by subtracting the 100 K data as the lattice contribution from the 50 K data below $T_N$ in figure 8(b). There are no discernible additional peaks; the spikes are caused by the thermal contraction of the lattice constant. The magnetic diffraction profile was calculated assuming a $q = 0$ AIAO-type magnetic order with a magnetic moment of $1\mu_B$ per Os, which contains some magnetic peaks with indices such as (2 2 0) and (1 1 3). Among them, the most intense (2 2 0) reflection, which is forbidden by crystal symmetry and is purely magnetic in origin, has an intensity equal to 0.5 percent that of the strongest (2 2 2) lattice reflection. A comparison of the experimental and simulated patterns reveals that the magnetic peaks are too small to observe within the experimental error margin in this experiment; the (2 2 0) peak would be almost completely obscured by noise. Notably, previous powder ND experiments did not resolve magnetic peaks for $Cd_2Os_2O_7$ [39], but they were resolved by a recent experiment in which an ordered magnetic moment of $0.59\mu_B$ per Os was estimated [27]. Due to the itinerant nature of $Hg_2Os_2O_7$, a smaller magnetic moment is expected, which would result in even smaller magnetic peaks. Additionally, it is noted that sample inhomogeneity broadens the magnetic peaks, posing an additional obstacle to their observation. Thus, while we were unable to detect magnetic reflections, this does not rule out the LRO. As a result, our ND experiments and those conducted previously [39] should not exclude the LRO, and the results may be consistent with itinerant magnetism with a reduced magnetic moment.

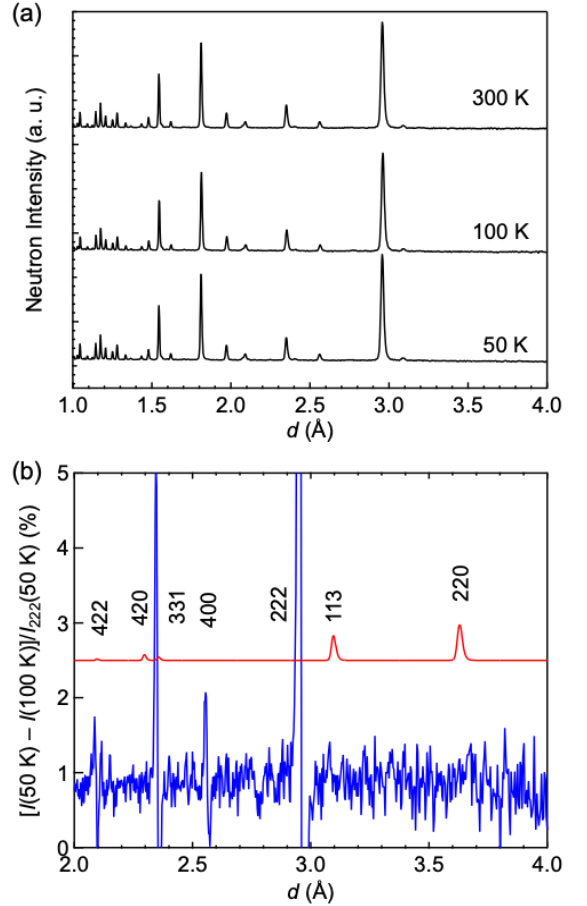

**Figure 8.** (a) Powder ND profiles at 300, 100, and 50 K. (b) Intensity difference between the 50 and 100 K data normalised to the (2 2 2) peak intensity $I_{222}$ at 50 K (bottom) and the corresponding simulated magnetic reflection profile assuming an ALAO-type order with a magnetic moment of $1\mu_B$ per Os (top), which has been shifted upward by 2.5 percent for clarity.

## 4. Discussion
### 4.1 $Cd_2Os_2O_7$ versus $Hg_2Os_2O_7$

To begin, we will compare the properties of the two compounds in figure 9. The temperature dependences of the magnetic susceptibility are alike: there are sharp cusps at $T_N$, followed by thermal hysteresis between the ZFC and FC curves. For $Cd_2Os_2O_7$, it is well established that an AIAO-type magnetic LRO sets in at $T_N$, whereas $\mu$SR experiments demonstrate a distinct LRO for $Hg_2Os_2O_7$. We hypothesise that $Hg_2Os_2O_7$ has a similar AIAO spin arrangement based on the observation of the robust ferromagnetism in common, which must originate from MDWs in the AIAO-type order. There is a significant difference in $T_N$: $T_N$ is much lower for $Hg_2Os_2O_7$.

CW magnetic susceptibilities are commonly observed above $T_N$, as in strongly correlated electron systems such as $V_2O_3$ [40]. Both compounds have very similar magnitudes and slopes (the two curves in figure 9 are plotted with the same scale on the right ordinate). Fitting the equation $\chi = \chi_0 + C/(T - \Theta)$ yields $T$-independent terms $\chi_0$ of $5.06(3) \times 10^{-4}$ [$4.9(1) \times 10^{-4}$] cm$^3$ mol$^{-1}$, Curie constants $C$ of 0.127(3) [0.122(9)] cm$^3$ K$^{-1}$ mol$^{-1}$ and Weiss temperatures $\Theta$ of –237(6) [– 131(15)] K for the Hg (Cd) compound. The Curie constants are used to calculate nearly equal effective magnetic moments $\mu_{\text{eff}}$ per Os of $0.71\mu_B$ ($0.70\mu_B$) using the equation $C/2 = (N_A/3k_B)\mu_{\text{eff}}^2$, where $N_A$ is Avogadro's number and $k_B$ is Boltzmann's constant. The nearly identical $\mu_{\text{eff}}$ values may reflect the similar electronic states of the paramagnetic phases. These values are significantly less than $3.87\mu_B$ for localised $d^3$ electrons with spin 3/2. Note that the $\mu_{\text{eff}}$ of the metallic phase of $V_2O_3$ is $2.37\mu_B$, which is comparable to the value of $2.83\mu_B$ for spin 1 [40]. The large reductions in osmates are due to the relatively low electron correlations and the large SOIs of the itinerant $5d$ electrons.

In contrast to the nearly identical $\mu_{\text{eff}}$ values, the magnitudes of the negative Weiss temperatures for the two osmates are quite different. Surprisingly, $|\Theta|$ does not scale with $T_N$. Moreover, the $|\Theta|$ of $Cd_2Os_2O_7$ is smaller than $T_N$. Due to geometrical frustration, it is frequently observed that $T_N$ is much smaller than $|\Theta|$ for insulating pyrochlore antiferromagnets. However, frustration may not be a critical factor in these itinerant magnets, and their Weiss temperatures do not appear to be a straightforward proxy for net magnetic interactions. According to spin fluctuation theory [41], the Curie constant and Weiss temperature of an itinerant magnet are not simply proportional to the magnitude and interactions of corresponding localised magnetic moments, respectively, but are dependent on the details of spin fluctuation modes [42].

Despite their similar magnetic susceptibilities, the resistivities below $T_N$ are quite different: $Cd_2Os_2O_7$ and $Hg_2Os_2O_7$ exhibit highly insulating and more metallic behaviours, respectively (figure 9). Notably, the $\rho$ values at room temperature are similar (~1 mΩ cm), whereas those at 2 K are four orders of magnitude different. Thus, for $Cd_2Os_2O_7$ and $Hg_2Os_2O_7$, phase transitions occur from similar paramagnetic semimetals with moderate electron correlations to an AIAO insulator and, most likely, to an AIAO semimetal at low temperatures, respectively.

*4.2 Electronic structures of paramagnetic phases*

The band structures of $5d$ pyrochlore oxides commonly possess $t_{2g}$ manifolds consisting of twelve bands near the Fermi level (four Os atoms per primitive cell), which are strongly hybridised with oxygen $2p$ states, and $e_g$ manifolds located much higher [12,13]. The $t_{2g}$ manifold is exactly half-filled for the present $Os^{5+}$ pyrochlores with the $5d^3$ electron configuration. However, unlike the MH case, the electron correlation $U$ cannot be large due to the large total bandwidth of ~3 eV and the multiband nature of the $5d$ orbital; unlike the $3d$ orbital, the extended $5d$ orbital should not result in a large on-site $U$. Indeed, it was estimated that $Cd_2Os_2O_7$ has an effective $U$ of 1.5 eV [15], which is less than the several eV $U$ of $3d$ electrons [36]. This must also be the case for $Hg_2Os_2O_7$ as well. On the other hand, the SOI is significantly large for a $5d$ electron with an energy of around 0.5 eV [1].

It is emphasized that the twelve bands comprising the $t_{2g}$ manifold of $Cd_2Os_2O_7$ are highly dispersive over a total of ~3 eV in the absence of the SOI, but split into two bundles, composed of six bands each in the presence of the SOI [12,13]. They are separated due to anti-band crossing; we refer to them as the lower and upper bundles, respectively, as $t_{2g}^{\text{low}}$ and $t_{2g}^{\text{up}}$. The separation may be enlarged by filling the $t_{2g}^{\text{low}}$ with more chemical bonding energy. However, because the separation is not complete in terms of energy, $t_{2g}^{\text{low}}$ is almost completely filled with a small number of holes remaining at the top, while $t_{2g}^{\text{up}}$ is almost completely empty with the same number of electrons at the bottom (figure 10). Their overlap is only about 0.12 eV. As a result, the compound transforms into a compensated metal with a variable carrier density that increases with increasing the overlap magnitude [12,13]. Note that this is a characteristic of the band structures of $5d$ pyrochlore oxides: for example, in $Cd_2Re_2O_7$ with the $5d^2$ electron configuration, the $t_{2g}$ manifold splits into $t_{2g}^{\text{low}}$ and $t_{2g}^{\text{up}}$, which are composed of four and eight bands, respectively, with $t_{2g}^{\text{low}}$ nearly filled [13].

It is critical to emphasize that the separation of $t_{2g}^{\text{low}}$ and $t_{2g}^{\text{up}}$, and thus their overlap, should depend on the SOI magnitude, as the gap created by anti-band crossing may increase with increasing SOI. When a small overlap is removed via a specific effect, the compound becomes insulating with an indirect gap opening without breaking symmetry. This is true for $Cd_2Os_2O_7$ but not for $Cd_2Re_2O_7$ due to the increased band overlap and decreased magnetic instability for the $d^2$ electron configuration.

*4.3 MITs in $Cd_2Os_2O_7$*

The MIT is one of the most fascinating properties of transition metal compounds and has been studied extensively for nearly a century [36,43,44]. In his textbook, Mott classified MITs into two categories: the band crossing transitions and MH transitions [36]. Typically, the band crossing transition occurs as a result of volume change under pressure that is not accompanied by magnetism. In contrast, the MH transition occurs when a paramagnetic metal (PM) is converted to an antiferromagnetic insulator (AFI) via $U$. Moreover, an antiferromagnetic metal (AFM) is frequently found between the PM and AFI, as is the case with vanadium-deficient $V_2O_3$ and $Ni(S,Se)_2$ [45,46]. Magnetic orders in AFMs are complex such as a spiral SDW order [47].

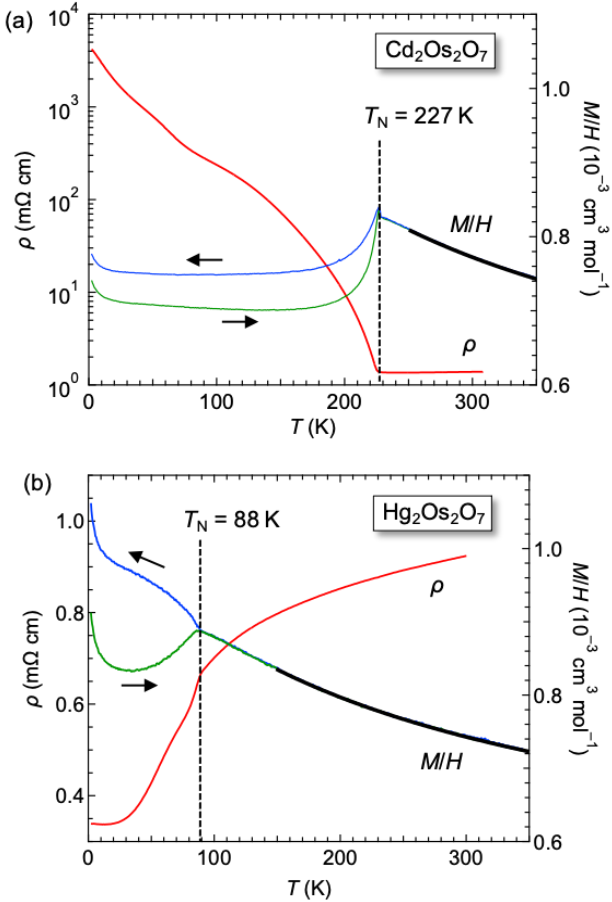

**Figure 9.** Comparison of the phase transitions in resistivity and magnetic susceptibility for (a) $Cd_2Os_2O_7$ and (b) $Hg_2Os_2O_7$ (crystal A, figure 3). The magnetic susceptibility of $Cd_2Os_2O_7$ was measured on a single crystal in a magnetic field of 2 T applied along the [1 1 1] direction [21]. The thick black lines at high temperatures corresponds to CW fits to the magnetic susceptibility data at 250–350 and 150–400 K, respectively.

The magnetic Lifshitz transition has been proposed for $Cd_2Os_2O_7$ [20,23,24]; it is not a Slater transition [48], as no Brillouin zone folding is induced by the $q = 0$ AIAO-type order. The Lifshitz transition is an electronic transition associated with a change in Fermi surface topology [49]; for example, a pressure-induced transition in $AuIn_2$ has been ascribed to a pocket-vanishing- or neck-collapsing-type Lifshitz transition [50]. Band crossing transitions, or MITs of the Bloch–Wilson-type [51], are a type of Lifshitz transition. What appears to distinguish the MIT of $Cd_2Os_2O_7$ from other Lifshitz transitions is the presence of magnetic order concomitantly.

$Cd_2Os_2O_7$'s AIAO-type order has been attributed to antiferromagnetic interactions induced by $U$ in the pyrochlore lattice and the local <1 1 1> anisotropy induced by SOIs [15,16,52,53]. As mentioned previously, $U$ in $Cd_2Os_2O_7$ is smaller than in $3d$ electron systems, but it is still quite large, as directly evidenced by the CW magnetic susceptibility of the PM phase. Nevertheless, the $U$ of ~1.5 eV predicted by the density functional theory [15] is exceptionally large for the $5d$ pyrochlore family; $U$ is almost negligible in $Cd_2Re_2O_7$ [54]. This moderate $U$ is caused by half-filling of the $t_{2g}$ manifold and can be ascribed to the sharp peak in the DOS at $E_F$, which originates from the nearly flat bands generated by SOI splitting of the $t_{2g}$ manifold [12,13]. Another factor that may contribute to the increased effective $U$ is the low carrier density, which results in a weak screening of Coulomb interactions.

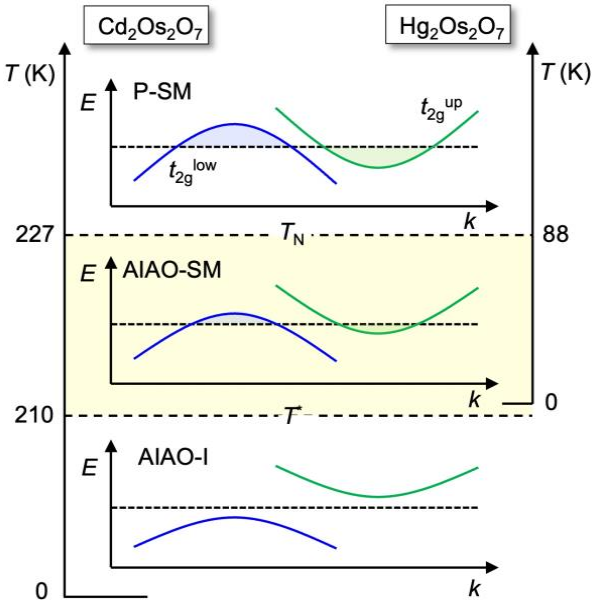

**Figure 10.** Schematic representations of the temperature evolutions of the semimetallic bands in $Cd_2Os_2O_7$ and $Hg_2Os_2O_7$. Commonly in the paramagnetic semimetal (P-SM) phases at high temperatures, the SOI devides the $t_{2g}$ manifold into nearly filled lower ($t_{2g}^{low}$) and nearly vacant upper ($t_{2g}^{up}$) bands with a small overlap. At $T_N$ (227 and 88 K for the Cd and Hg compounds, respectively), a magnetic LRO occurs in a second-order manner to remove the magnetic instability associated with moderate electron correlations. This results in a band shift or flattening, which eliminates the overlap via enhanced SOIs and effective $U$. As a result, an all-in–all-out semimetal (AIAO-SM) state is stabilised. Further metal–insulator crossover occurs at $T^* \sim$ 210 K for $Cd_2Os_2O_7$, via an indirect gap opening towards an AIAO insulator (AIAO-I), whereas the AIAO-SM survives to $T = 0$ for $Hg_2Os_2O_7$ due to its lower magnetic instability and greater band overlap in the original P-SM. The two compounds are separated by a delicate balance of electron correlations and SOIs on either side of the metal–insulator boundary.

The magnetic instability associated with $U$ in paramagnetic semimetals (P-SM) results in the AIAO-type order below $T_N$ in $Cd_2Os_2O_7$. Once the AIAO-type order is established, spin polarisation can effectively enhance the SOI effect, which may increase the separation between the $t_{2g}^{low}$ and $t_{2g}^{up}$ bands. Moreover, when the MIT is approached, the effective $U$, and thus the magnetic instability, can be increased by decreasing the carrier density. On the other hand, approaching a localised state may cause band flattening. Then, as the AIAO-SM cools, the band overlap gradually decreases, as schematically illustrated in figure 10. At $T^*$, an indirect gap forms, resulting in the AIAO insulator (AIAO-I) as the ground state. However, because thermally excited carriers are present, the transition appears to be a crossover. This is the scenario for the magnetic Lifshitz MIT. Note that both $U$ and SOI play a role in this unique MIT. In addition, similarly to the MH transition, the MIT occurs via an AFM phase (AIAO-SM) at $T_N$–$T^*$.

*4.4 Phase transition in $Hg_2Os_2O_7$*

In $Hg_2Os_2O_7$, a very similar transition should occur. As previously stated [20], $Hg_2Os_2O_7$ has a band structure similar to that of $Cd_2Os_2O_7$ but with a larger band overlap of approximately 0.2 eV due to the influence of the Hg $6s$ band located at lower energies than the Cd $5s$ band at high energies above $E_F$: these bands tend to push down $t_{2g}^{up}$ to increase the overlap with $t_{2g}^{low}$. The increased overlap results in a higher carrier density, which matches our Hall coefficient data. Moreover, the more dispersive bands near $E_F$ result in a smaller DOS peak. Therefore, one would expect a smaller $U$ and thus less magnetic instability in $Hg_2Os_2O_7$ than in $Cd_2Os_2O_7$, which explains $Hg_2Os_2O_7$'s lower $T_N$.

Due to the reduced magnetic instability and increased band overlap, the band shift induced by the AIAO-type order decreases and eventually becomes insufficient to completely eliminate the band overlap, resulting in an AIAO-SM ground state, as illustrated in figure 10. One would anticipate an itinerant magnetic order, such as an AIAO-SDW order, to have more reduced magnetic moments on average than the more localised order found in the AIAO-I of $Cd_2Os_2O_7$, as in the MH type AFM. Recent $\mu$SR analysis appears to support such an AIAO-SDW order [38]. Thus, the MIT of $Cd_2Os_2O_7$ and the metal–metal transition of $Hg_2Os_2O_7$ can be interpreted reasonably in terms of the magnetic Lifshitz mechanism; however, this terminology may not be directly applicable to the phase transition in $Hg_2Os_2O_7$, as the Fermi surface topology does not change.

A remaining issue unresolved is the observed reduction in resistivity near $T_N$ for $Hg_2Os_2O_7$ (figure 9). For $Cd_2Os_2O_7$, the resistivity increases sharpy below $T_N$, which has been attributed to a decrease in carrier density with decreasing band overlap upon cooling in the intermediate AIAO-SM [20]; a true indirect gap opens below $T^*$, where the corresponding change in resistivity is undetectable due to the presence of thermally excited carriers across a tiny gap [23]. If $Hg_2Os_2O_7$ has a similar band shift, its resistivity should also initially increase just below $T_N$, rather than decrease, and then decrease upon further cooling in the AIAO-SM state.

In a previous μSR study on $Cd_2Os_2O_7$, Koda et al. demonstrated conducting small polarons produced by a strong spin–phonon coupling below $T^*$ [30]. This implies that understanding the transport properties of the AIAO-SM phase at high temperatures up to $T_N$ requires consideration of magnetoelastic coupling. As long as the Lifshitz transition occurs concomitantly with $U$ becoming dominant at low temperatures, the carrier density around $T_N$ will exhibit only a slight decrease. Thus, the increase in $\rho(T)$ below $T_N$ in $Cd_2Os_2O_7$ is attributed to a decrease in mobility by the formation of a polaronic state via magnetoelastic coupling in the AIAO-SM phase. This is confirmed by the strong concordance in the temperature dependences of the plasma frequency [23] and the spin-coupled phonons frequency [28]. By contrast, the magnetoelastic coupling in $Hg_2Os_2O_7$ may be small due to the larger band overlap and associated higher carrier density. Thus, as magnetic scattering decreases in the AIAO-SM phase, the resistivity simply decreases below $T_N$ in $Hg_2Os_2O_7$.

## 5. Conclusions and remarks

We investigated and compared the phase transition of the $5d$ pyrochlore oxide $Hg_2Os_2O_7$ to that of the isoelectronic $Cd_2Os_2O_7$. In contrast to the transition to an AIAO-SM below 227 K and then to an AIAO-I below $T^*$ in $Cd_2Os_2O_7$, it was discovered that $Hg_2Os_2O_7$ exhibits a magnetic transition, most likely to an AIAO-type order, below $T_N = 88$ K while retaining its metallic conductivity. The difference is ascribed to the $Hg_2Os_2O_7$'s lower magnetic instability (electron correlations) and a greater overlap of the hole and electron bands than $Cd_2Os_2O_7$'s. Both electron correlations and SOIs are suggested to play a significant role in these magnetically induced MIT and metal–metal transitions.

$Cd_2Os_2O_7$ and $Hg_2Os_2O_7$ exist on either side of the MI boundary. On the metal side, one expects interesting phenomena, particularly for $Hg_2Os_2O_7$: for example, a magnetic fluctuation may induce an exotic superconductivity. However, we did not observe superconductivity or other unusual phenomena above 2 K, most likely due to poor sample quality. We are currently attempting to improve the quality of the sample and will return to this intriguing possibility in the future. Moreover, a quantum critical point between AIAO-SM and AIAO-I may be reached in a solid solution of $(Cd, Hg)_2Os_2O_7$ or under pressure, in which gap will close at $T = 0$, resulting in increased excitonic instability [12]. Osmium pyrochlores continue to be fascinating and will garner increased attention in the future.

**Acknowledgments**
We would like to express our gratitude to D. Nishio-Hamane for the photographs in figure 1. The μSR experiments were conducted under the auspices of TRIUMF Exp. No. M1711, and the authors would like to thank the TRIUMF staff for their technical assistance throughout the experiments. This work was supported by the Japan Society for the Promotion of Science (JSPS) through KAKENHI Grant Numbers JP18H01169 and JP20H05150 (Quantum Liquid Crystals). The neutron scattering experiment received approval from the Neutron Scattering Program Advisory Committee of IMSS, KEK (Proposal No. 2014S06).